\documentclass[aps,prd,nofootinbib,amsmath,amssymb,superscriptaddress,showpacs,showkeys,twocolumn,10pt] {revtex4}

\usepackage{graphicx}
\usepackage{dcolumn}
\usepackage{bm}
\usepackage{amssymb}
\usepackage{latexsym}
\usepackage{booktabs}
\usepackage{amsmath}
\usepackage{multirow}
\usepackage[colorlinks=true, linkcolor=red, citecolor=blue]{hyperref}
\usepackage{color}
\newcommand{\be}{\begin{equation}}
\newcommand{\ee}{\end{equation}}
\newcommand{\bq}{\begin{eqnarray}}
\newcommand{\eq}{\end{eqnarray}}

\begin{document}

\title{Redshift drift constraints on  holographic dark energy }

\author{Dong-Ze He}
\affiliation{Department of Physics, College of Sciences, Northeastern University, Shenyang
110004, China}
\author{Jing-Fei Zhang}
\email{jfzhang@mail.neu.edu.cn} 
\affiliation{Department of Physics, College of Sciences, Northeastern University, Shenyang
110004, China}
\author{Xin Zhang}
\affiliation{Department of Physics, College of Sciences,
Northeastern University, Shenyang 110004, China}
\affiliation{Center for High Energy Physics, Peking University, Beijing 100080, China}

\begin{abstract}

The Sandage-Loeb (SL) test is a promising method for probing dark energy because it measures the redshift drift in the spectra of Lyman-$\alpha$ forest of distant quasars, covering the ``redshift desert" of $2\lesssim z\lesssim5$, which is not covered by existing cosmological observations. Therefore, it could provide an important supplement to  current cosmological observations. In this paper, we explore the impact of SL test on the precision of cosmological constraints for two typical holographic dark energy models, i.e., the original holographic dark energy (HDE) model and the Ricci holographic dark energy (RDE) model. To avoid data inconsistency, we use the best-fit models based on current combined observational data as the fiducial models to simulate 30 mock SL test data. The results show that SL test can effectively break the existing strong degeneracy between the  present-day matter density $\Omega_{m0}$ and the Hubble constant $H_0$ in other cosmological observations. For the considered two typical dark energy models, not only can a 30-year observation of SL test improve the constraint precision of $\Omega_{m0}$ and $h$ dramatically, but can also enhance the constraint precision of the model parameters $c$ and $\alpha$ significantly.
\end{abstract}


\maketitle

\section{Introduction}
\label{sec:intro}
At the end of last century, the type Ia supernovae observations discovered that our universe is undergoing an accelerating expansion \cite{Riess:1998cb,Perlmutter:1998np}. In order to explain this apparently counterintuitive behavior of the universe, a mysterious energy component,
dubbed ``dark energy''~(DE), is usually assumed to exist and dominate the evolution of current universe. However, other
than the fact that it is almost uniformly distributed, gravitionally repulsive, and contributes about 70\% of the total
energy in the universe, people actually know little about the nature of DE. In spite of this, cosmologists still have already proposed
numerous DE models, making attempts to uncover its mystery.

On the other hand, if one wishes to place more comprehensive cosmological constraints on a underlying cosmological model and then precisely acknowledge the geometry and matter contents of the universe, it should be necessary to measure the expansion rate of universe at different redshifts. Among all the known datasets, cosmic background microwave anisotropies (CMB) measurements probe the rate of expansion at the redshift $z\sim1100$, while for much lower redshift~($z<2$), the expansion history measurements could depend on weak lensing, baryon acoustic oscillation~(BAO), type Ia supernovae~(SN) and so forth. However, up to now, the redshift range between $z\sim2$ to 1100, regarded as ``redshift desert", is still a blank area for which the existing dark energy probes are unable to provide useful
information about the expansion history of our universe.
Therefore, the redshift drift data in the ``redshift desert" of $2\lesssim z\lesssim5$ will provide an important supplement to the current observational data
and play a more significant role in future parameter constraints.
Redshift drift observation is a purely geometric measurement of the expansion of the universe, which was originally proposed by Sandage to directly measure the temporal variation of the redshift of extra-galactic sources in 1962 \cite{sandage}, and then improved by Loeb in 1998 who suggested the possibility of measuring the redshift drift by decades-long observation of the redshift variation of distant quasars (QSOs) Lyman-$\alpha$ absorption lines \cite{loeb}. Thus, the method of redshift drift measurement is also referred to as the ``Sandage-Loeb" test.

The Sandage-Loeb (SL) test is a unique method to directly measure the cosmic expansion rate in the ``redshift desert" of $2\lesssim z\lesssim5$, which is never covered by any other existing dark energy probes. Combining the SL test data from this high-redshift range with other data from low-redshift region, such as the 
SN, the 
BAO and the like, will definitely lead to significant impact on the dark energy constraints. The scheduled 39-meter European Extremely Large Telescope (E-ELT) equipped with a high-resolution spectrograph called the Cosmic Dynamics Experiment (CODEX) is designed to collect such SL test signals. A great deal of work has been done on the effect of the SL test on cosmological parameter estimation \cite{sl1,sl2,sl3,sl4,sl5,sl6,sl7}, some of which improperly assumed 240 or 150 observed QSOs in the simulations. Nevertheless, on the strength of an extensive Monte Carlo simulation, using a telescope with a spectrograph like CODEX, only about 30 QSOs will be bright enough and/or lie at a high enough redshift for the actual observation. Furthermore, as is known to all, in most existing papers about the SL test, the best-fit $\Lambda$ cold dark matter ($\Lambda$CDM) model is usually chosen as the fiducial model in simulating the mock SL test data. In this way, when these simulated SL test data are combined with other data to constrain some dynamical dark energy models (or modified gravity models), tension may exist among the combined data, leading to an inappropriate joint constraint.

In our previous works \cite{msl1,msl2,msl3,msl4}, we quantified the impact of future redshift-drift measurement on parameter estimation for different dark energy models. In order to correctly quantify the impact of the future SL test data on dark energy constraints, producing the simulated SL test data consistent with other actual observations is extremely significant and indispensable. Here, we have to point out that the SL test data alone cannot tightly constrain dark energy models owing to the lack of low-redshift data. For this reason, the combination of simulated SL test data with currently available actual data covering the low-redshift region is supposed to be very necessary for the constraints on dark energy. On the other hand, when we combine the SL test data with other current observational data, the existing parameter degeneracies in current observations will be broken effectively, with the precision of parameter estimation in the widely studied dark energy models improved greatly at the same time \cite{msl1,msl2,msl3,msl4}. And, in order to eliminate the potential inconsistencies between the current data and simulated future SL data, we decide to choose the best-fitting dark energy models as the fiducial models to produce the simulated future data.

Among all the existing dark energy models, the holographic dark energy model, which is a dynamical DE model based on the holographic principle of quantum gravity, is a very competitive candidate for DE. Based
on the effective quantum field theory, Cohen et al. \cite{Cohen} pointed out that, if gravity is considered, the total energy of a system with size $L$ should not exceed the mass of a black hole with the same size, i.e.,
$L^{3}\rho_{de}\lesssim LM_{pl}^{2}$. This energy bound leads to the density of holohraphic dark energy,
\begin{equation}
   \rho_{de}=3c^{2} M_{pl}^{2}L^{-2},
\end{equation}
where $c$ is a dimensionless parameter characterizing some uncertainties in the effective quantum field theory, $M_{pl}$ is the reduced Planck mass defined by $M_{pl}^2=(8\pi G)^{-1}$, and $L$
is the infrared~(IR) cutoff in the theory. Li \cite{Li} suggested that the IR cutoff $L$ should be given by the future event horizon of the universe. This yields the original holographic dark energy model (see \cite{Wang:2013zca,Cui:2015oda,Guo:2015gpa,Xu:2016grp,Feng:2016djj,Zhang:2015uhk,Wang:2016tsz} for recent constraints). Furthermore, Gao et al. \cite{CJGao} proposed to consider the average radius of the Ricci scalar curvature as the IR cutoff, and this model is called the holographic Ricci dark energy model (see also \cite{Zhang:2009un}). For convenience, hereafter we will call them HDE and RDE, respectively.
Recently, Geng et al. \cite{msl1,msl2,msl3,msl4} employed the simulated Sandage-Loeb test data to explore many different kinds of dark energy models. In these analyses, the two popular and competitive models, namely, the HDE model and the RDE model, are absent.
Thus, as a further step along this line, in this paper we will provide such an analysis to make the analysis of the constraining power of the future SL test on dark energy more general and complete. 

The organization of this paper is as follows. In Sec. \ref{sec:model}, we will briefly review the holographic dark energy models. In Sec. \ref{sec:cosmol}, we will present the observational data used in this work, as well as the basic introduction of the SL test. In Sec. \ref{sec:resul}, we will show the results of the cosmological constraints, and quantify the improvement in the parameter constraints from the SL test. Conclusion will be given in Sec. \ref{sec:conclu}.

\begin{figure}
\includegraphics[width=9cm]{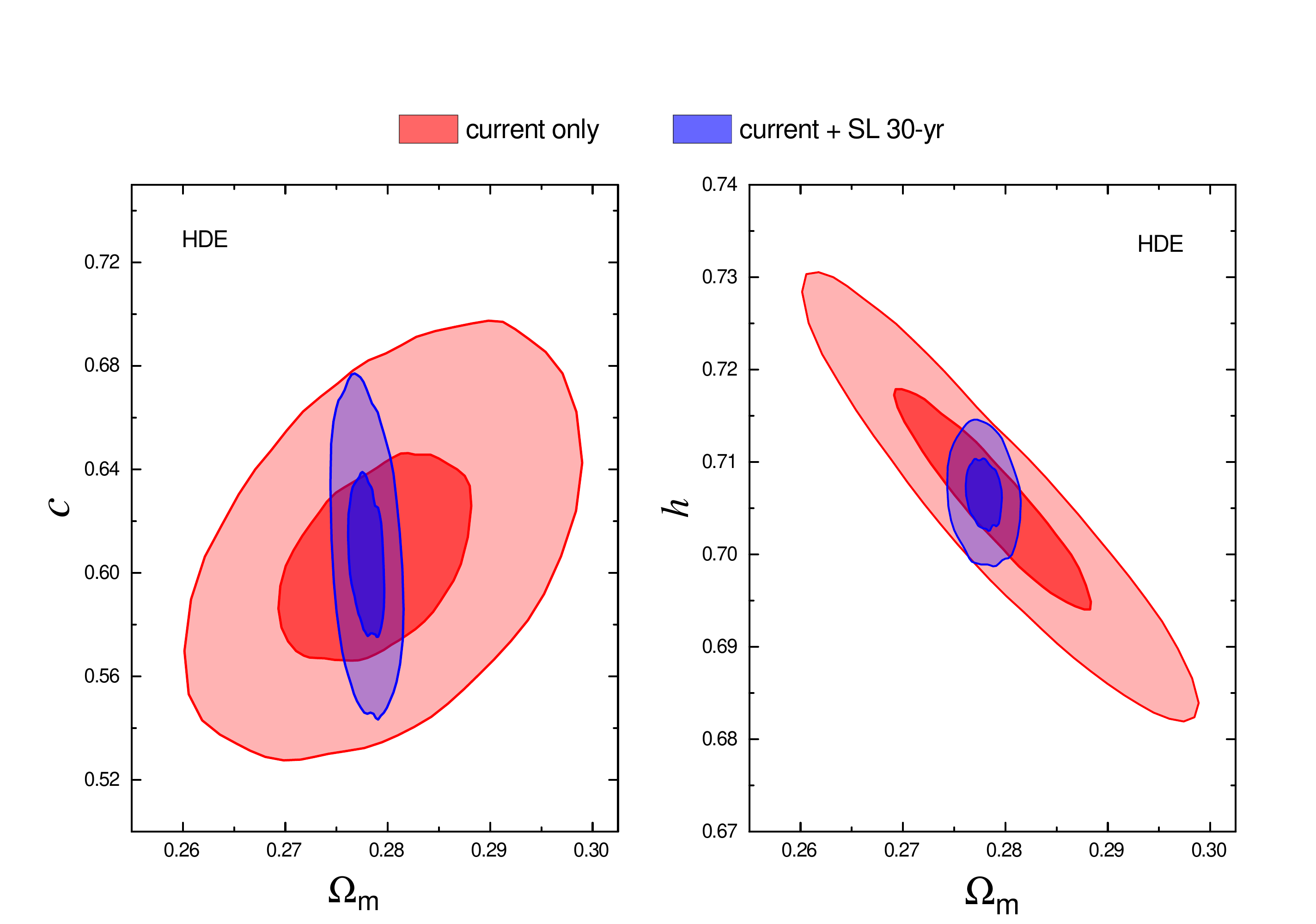}
\caption{\label{fig1}Constraints (68.3\% and 95.4\% CL) in the $\Omega_{m0}$-$c$ plane and in the $\Omega_{m0}$-$h$ plane for HDE model with current only and current+SL 30-year data. }
\end{figure}

\begin{figure}
\begin{center}
\includegraphics[width=9cm]{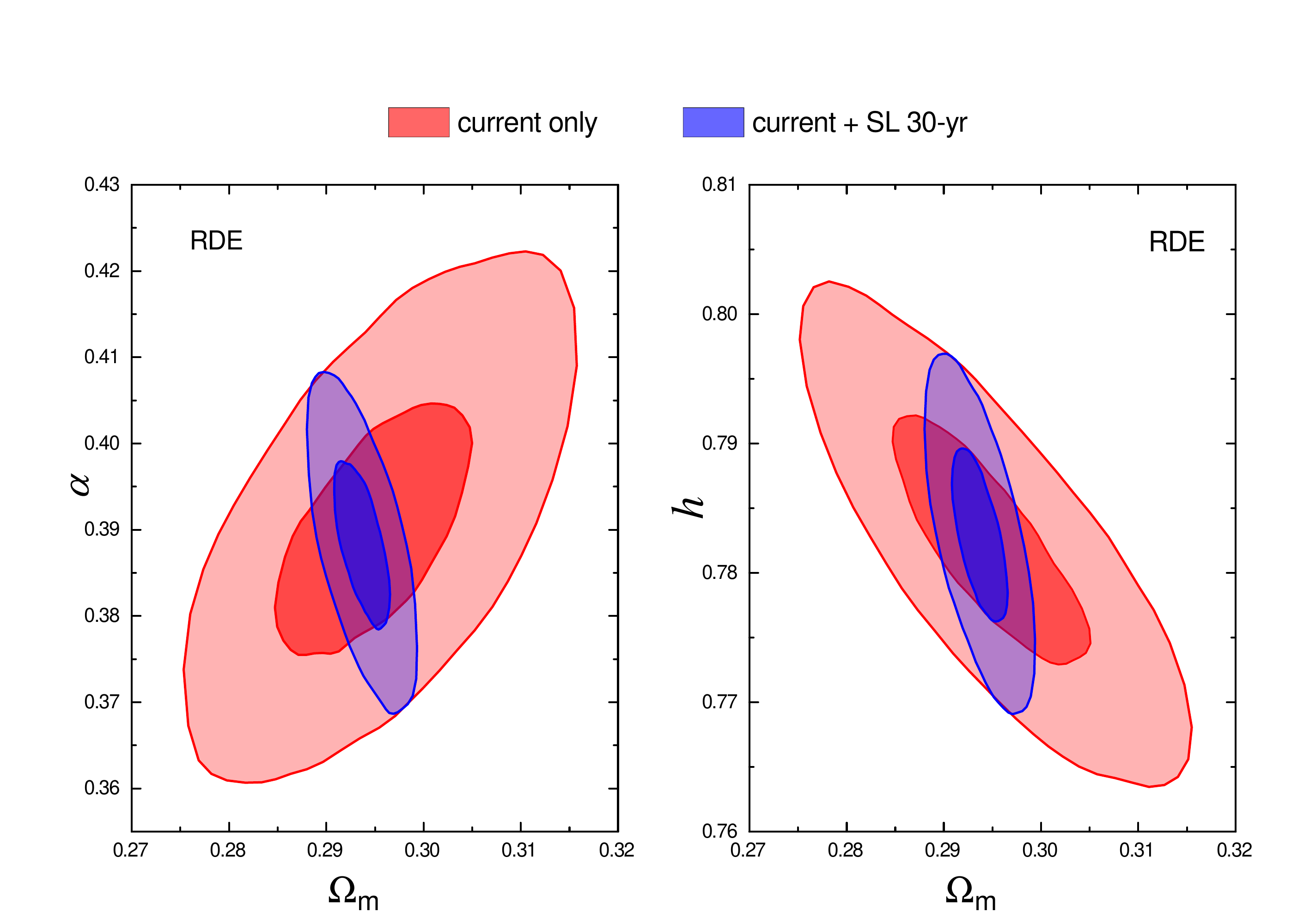}
\end{center}
\caption{\label{fig2}Constraints (68.3\% and 95.4\% CL) in the $\Omega_{m0}$-$\alpha$ plane and in the $\Omega_{m0}$-$h$ plane for RDE model with current only and current+SL 30-year data. }

\end{figure}

\section{ Models}\label{sec:model}
In this section, we shall briefly review the original holographic dark energy model and the Ricci dark energy model. In fact, these two models both belong to the holographic scenario of dark energy. 

For a spatially flat (the assumption of flatness is motivated by the inflation scenario and, actually, the current observations strongly favor a flat universe) Friedmann-Robertson-Walker (FRW) universe with matter component $\rho_m$ and dark energy component $\rho_{de}$, the Friedmann equation reads
\begin{equation}\label{2.1}
3M_{pl}^{2}H^{2}=\rho_m+\rho_{de},
\end{equation}
or equivalently,
\begin{equation}\label{2.2}
E(z)\equiv\frac{H(z)}{H_{0}}=\bigg(\frac{\Omega_{m0}(1+z)^{3}}{1-\Omega_{de}}\bigg)^{1/2},
\end{equation}
where $H\equiv\dot{a}/a$ is the Hubble parameter, $\Omega_{m0}$ is the present-day fractional matter density, and $\Omega_{de}\equiv\frac{\rho_{de}}{\rho_{c}}=\frac{\rho_{de}}{3M_{pl}^{2}H^{2}}$ is the fractional dark energy density. The energy conservations of matter and dark energy give
\begin{equation}\label{2.3}
\dot{\rho}_{m}+3H\rho_{m}=0,
\end{equation}
\begin{equation}\label{2.4}
\dot{\rho}_{de}+3H(1+w_{de})\rho_{de}=0,
\end{equation}
where the overdot denotes the derivative with respect to the cosmic time $t$, and $w_{de}$ is the EOS of DE.

\subsection{The HDE model}\label{sec:HDE}

For this model, the IR cutoff is chosen as the future event horizon of the universe,
\begin{equation}\label{2.5}
L=a\int^{\infty}_{t}\frac{dt}{a}=a\int^{\infty}_{a}\frac{da}{Ha^{2}}.
\end{equation}
Taking derivative for $\rho_{de}=3c^{2}M_{pl}^{2}L^{-2}$ with respect to $x=\ln a$ and making use of Eq.~(\ref{2.5}), we get
\begin{equation}\label{2.6}
\rho_{de}'\equiv\frac{d\rho_{de}}{dx}=2\rho_{de}\bigg(\frac{\sqrt{\Omega_{de}}}{c}-1\bigg).
\end{equation}
Combining Eqs. (\ref{2.4}) and (\ref{2.6}), we obtain the EOS for HDE,
\begin{equation}\label{2.7}
w_{de}=-\frac{1}{3}-\frac{2}{3c}\sqrt{\Omega_{de}}.
\end{equation}
Directly taking derivative for $\Omega_{de}=c^{2}/(H^{2}L^{2})$, and using Eq. (\ref{2.5}), we get
\begin{equation}\label{2.8}
\Omega_{de}'=2\Omega_{de}\bigg(\epsilon-1+\frac{\sqrt{\Omega_{de}}}{c}\bigg),
\end{equation}
where $\epsilon\equiv-\dot{H}/H^{2}=-H'/H$. From Eqs. (\ref{2.1}), (\ref{2.3}), (\ref{2.4}), and (\ref{2.7}), we have
\begin{equation}\label{2.9}
\epsilon=\frac{3}{2}(1+w_{de}\Omega_{de})=\frac{3}{2}-\frac{\Omega_{de}}{2}-\frac{\Omega_{de}^{3/2}}{c},
\end{equation}
for this case. So, we have the equation of motion, a differential equation, for $\Omega_{de}$,
\begin{equation}\label{2.10}
\Omega_{de}'=\Omega_{de}(1-\Omega_{de})\bigg(1+\frac{2}{c}\sqrt{\Omega_{de}}\bigg).
\end{equation}
Since $\frac{d}{dx}=-(1+z)\frac{d}{dz}$, we get
\begin{equation}\label{2.11}
\frac{d\Omega_{de}}{dz}=-(1+z)^{-1}\Omega_{de}(1-\Omega_{de})\bigg(1+\frac{2}{c}\sqrt{\Omega_{de}}\bigg).
\end{equation}
Solving numerically Eq.~(\ref{2.11}) and substituting the corresponding results into Eq.~(\ref{2.2}), the function $E(z)$ can be obtained. It should be mentioned that there are two model parameters, $\Omega_{m0}$ and $c$, in the HDE model.

\subsection{The RDE model}\label{sec:RDE}

For a spatially flat FRW universe, the Ricci scalar is
\begin{equation}\label{2.12}
\mathcal {R}=-6 \bigg(\dot{H}+2H^{2}\bigg).
\end{equation}
 As suggested by Gao et al.~\cite{CJGao}, the energy density of dark energy is given by
\begin{equation}\label{2.13}
\rho_{de}=\frac{3\alpha}{8\pi G}(\dot{H}+2H^{2})=-\frac{\alpha}{16\pi G}\mathcal{R},
\end{equation}
where $\alpha$ is a positive numerical constant to be determined by observations. Comparing to $\rho_{de}=3c^{2}M_{pl}^{2}L^{-2}$, it is seen that if we identify the IR cutoff $L$ with $-{\cal R}/6$, we have $\alpha=c^{2}$. As pointed out by Cai et al. \cite{Cai}, the RDE can be viewed as originated from taking the causal connection scale as the IR cutoff in the holographic setting. Now the Friedmann equation can be written as
\begin{equation}\label{2.14}
H^{2}=\frac{8\pi G}{3}\rho_{m0}e^{-3x}+\alpha\bigg(\frac{1}{2}\frac{dH^{2}}{dx}+2H^{2}\bigg),
\end{equation}
and this equation can be further rewritten as
\begin{equation}\label{2.15}
E^{2}=\Omega_{m0}e^{-3x}+\alpha\bigg(\frac{1}{2}\frac{dE^{2}}{dx}+2E^{2}\bigg),
\end{equation}
where $E\equiv H/H_{0}$. Solving this equation and using the initial condition $E_{0}=E(t_{0})=1$, we have
\begin{equation}\label{2.16}
E(z)=\bigg(\frac{2\Omega_{m0}}{2-\alpha}(1+z)^{3}+\bigg(1-\frac{2\Omega_{m0}}{2-\alpha}\bigg)(1+z)^{(4-\frac{2}{\alpha})}\bigg)^{1/2}.
\end{equation}
There are also two model parameters, $\Omega_{m0}$ and $\alpha$, in the RDE model.

\section{ methodology}\label{sec:cosmol}

First of all, we compactly introduce the current observational data utilized in this paper. In our analysis, for current data, the most typical and commonly quoted geometric measurements are chosen, i.e., the type Ia supernovae, the cosmic microwave background, the baryon acoustic oscillation, and the direct measurement of the Hubble constant $\rm H_0$. In fact, the combination of SN, BAO, CMB, and $\rm H_0$ is the most popular data intergration in parameter estimation studies of dark energy models. For the SN data, the SNLS compilation with a sample of 472 SNe \cite{snls3} is used. For the BAO data, we consider the $r_s/D_V(z)$ measurements from 6dFGS ($z=0.1$) \cite{6dF}, SDSS-DR7 ($z=0.35$) \cite{DR7}, SDSS-DR9 ($z=0.57$) \cite{DR9}, and WiggleZ ($z=0.44$, 0.60, and 0.73) \cite{WiggleZ} surveys. As dark energy only affects the CMB through the comoving angular diamater distance to the decoupling epoch (and the late-time ISW effect), the distance information given by the CMB distance prior is sufficient for the joint geometric constraint on dark energy. Hence, with regard to the CMB data, we adopt the Planck distance prior in Ref. \cite{WW}. Apart from that, the direct measurement result of $H_0$ in the light of the cosmic distance ladder from the Hubble Space Telescope, $H_0=73.8 \pm 2.4$ km s$^{-1}$ Mpc$^{-1}$ \cite{Riess2011}, is adopted in this work as well.

Secondly, when it comes to SL test, it is necessary to briefly review the basics of this method. This method is just to directly measure the redshift variation of quasar Lyman-$\alpha$ absorption lines. In the redshift drift measurement, the redshift variation is defined as the spectroscopic velocity shift \cite{Liske}
\begin{equation}\label{eq1}
\ \Delta v \equiv \frac{\Delta z}{1+z}=H_0\Delta t_o\bigg[1-\frac{E(z)}{1+z}\bigg],
\end{equation}
where $\Delta t_o$ is the time interval of the observation.
$E(z)=H(z)/H_0$ is determined by specific dark energy models.

According to a Monte Carlo simulation, the uncertainty of $\Delta v$ can be expressed as
\begin{equation}\label{eq2}
\sigma_{\Delta v}=1.35
\bigg(\frac{S/N}{2370}\bigg)^{-1}\bigg(\frac{N_{\mathrm{QSO}}}{30}\bigg)^{-1/2}\bigg(
\frac{1+z_{\mathrm{QSO}}}{5}\bigg)^{x}~\mathrm{cm}~\mathrm{s}^{-1},
\end{equation}
where $S/N$ is the signal-to-noise ratio defined per 0.0125 $\mathring{\rm A}$ pixel, and the last exponent is $x=-1.7$ for $2<z<4$ and $x=-0.9$ for $z>4$. $N_{\mathrm{QSO}}$ is the number of observed QSOs, and $z_{\mathrm{QSO}}$ denotes their redshift. We simulate $N_{\mathrm{QSO}}=30$ SL test data uniformly distributed over six redshift bins of $z_{\mathrm{QSO}}\in [2,~5]$ and typically apply $\Delta t_o = 30$ year in our analysis.

With respect to the simulation of the SL test data, the dark energy models will be constrained with the utilization of SN+BAO+CMB+$\rm H_0$ data. Actually, we can get the central values of the mock data by substituting the obtained best-fit parameters in the fit to the current data into Eq. (\ref{eq1}). The error bars can be computed from Eq. (\ref{eq2}) with $S/N$ = 3000.

Our procedure is as follows. Dark energy models are primarily constrained by using the current joint SN+BAO+CMB+$\rm H_0$ data, and then the best-fit dark energy models are selected to be the fiducial models in producing the simulated SL test data. Subsequently, the dark energy models will be constrained for a second time, using the simulated SL test data combined with the existing data, and the improvement in the parameter estimation of SL test will also be quantified.

\begin{table*}
\renewcommand{\arraystretch}{1.2}
\begin{tabular}{cccccc}
\hline\hline
 &\multicolumn{2}{c}{current only}&&\multicolumn{2}{c}{current+SL 30-year}\\
           \cline{2-3}\cline{5-6}
Parameter  & HDE & RDE&& HDE &RDE \\
\hline
$\Omega_bh^2$&$0.0224^{+0.0003}_{-0.0003}$&$0.0223^{+0.0003}_{-0.0003}$&&$0.0224^{+0.0003}_{-0.0002}$&$0.0223^{+0.0003}_{-0.0003}$\\
$\Omega_ch^2$&$0.1166^{+0.0019}_{-0.0020}$&$0.1576^{+0.0041}_{-0.0038}$&&$0.1165^{+0.0016}_{-0.0020}$&$0.1577^{+0.0026}_{-0.0027}$\\
$c$&$0.6022^{+0.0451}_{-0.0375}$&--&&$0.6013^{+0.0390}_{-0.0277}$&--\\
$\alpha$&--&$0.3881^{+0.0169}_{-0.0136}$&&--&$0.3887^{+0.0096}_{-0.0104}$\\
$\Omega_m$&$0.2790^{+0.0097}_{-0.0101}$&$0.2943^{+0.0112}_{-0.0101}$&&$0.2780^{+0.0018}_{-0.0019}$&$0.2935^{+0.0033}_{-0.0027}$\\
$h$&$0.7059^{+0.0122}_{-0.0121}$&$0.7820^{+0.0106}_{-0.0094}$&&$0.7069^{+0.0036}_{-0.0046}$&$0.7833^{+0.0064}_{-0.0071}$\\
\hline
\hline

\end{tabular}
\caption{\label{tab1}Fitting results for the RDE and HDE models using the current only and current+SL 30-year data. Here we quote $\pm1\sigma$ errors.}
\end{table*}

\section{results and discussion}\label{sec:resul}

We constrain the HDE and RDE models by using the current data (current only) and the combination of current data and the 30-year SL test data (current+SL 30-year). The detailed fit results are given in Table~\ref{tab1}, with the $\pm1\sigma$ errors quoted. Hereafter, ``current" denotes the current SN+BAO+CMB+$\rm H_0$ data combination for convenience. As can be seen from this table, when the SL 30-year data are combined, almost all the constraint results would be improved significantly.

\begin{table*}
\renewcommand{\arraystretch}{1.2}
\begin{tabular*}{7cm}{@{\extracolsep{\fill}}cccccc}
\hline\hline
&\multicolumn{2}{c}{current only}&&\multicolumn{2}{c}{current+SL 30-year}\\
           \cline{2-3}\cline{5-6}
Error  & HDE & RDE&& HDE &RDE \\
\hline
$\sigma(c)$&$0.0415$&--&&$0.0338$&--\\
$\sigma(\alpha)$&--&0.0153&&--&$0.0101$\\
$\sigma(\Omega_m)$&0.0099&0.0107&&0.0018&0.0030\\
$\sigma(h)$&0.0122&0.0100&&0.0041&0.0068\\ \hline\hline
\end{tabular*}
\caption{\label{tab2}Errors of parameters in the HDE and RDE models for the fits to current only and current+SL 30-yr data.}
\end{table*}

\begin{table*}
\renewcommand{\arraystretch}{1.2}
\begin{tabular*}{7cm}{@{\extracolsep{\fill}}cccccc}
\hline\hline
&\multicolumn{2}{c}{current only}&&\multicolumn{2}{c}{current+SL 30-year}\\
           \cline{2-3}\cline{5-6}
Precision  & HDE & RDE&& HDE &RDE \\
\hline
$\varepsilon(c)$&$6.89\%$&--&&$5.62\%$&--\\
$\varepsilon(\alpha)$&--&$3.94\%$&&--&$2.60\%$\\
$\varepsilon(\Omega_m)$&$3.55\%$&$3.64\%$&&$0.65\%$&$1.02\%$\\
$\varepsilon(h)$&$1.73\%$&$1.28\%$&&$0.58\%$&$0.87\%$\\ \hline\hline
\end{tabular*}
\caption{\label{tab3}Constraints precisions of parameters in the HDE and RDE models for the fits to current only and current+SL 30-yr data.}
\end{table*}

With the purpose of observing the improvements of parameter constraints from the SL test simulated data visually, we show the joint constraint results in Figures~\ref{fig1} and~\ref{fig2}. In Figure~\ref{fig1}, we show the 68.3\% and 95.4\% CL posterior distribution contours in the $\Omega_m-h$ plane and $\Omega_m-c$ plane for the holographic dark energy model. The current only and the current+SL 30-year results are presented in red and blue, respectively. In Figure~\ref{fig2}, we present the joint constraints on the Ricci dark energy model (68.3\% and 95.4\% CL) in the $\Omega_m-h$ and $\Omega_m-\alpha$ planes, with the current only constraint shown in red while the current+SL 30-year constraint exhibited in blue. From these figures, we clearly find that the degeneracy directions are evidently changed by adding the SL 30-year data.

In order to quantify the improvements, we list the errors of parameters in the HDE and RDE models for the fits to the current data and the current+SL 30-year data, in Table~\ref{tab2}. Because of the fact that the fit results are not in the form of totally normal distributions, we define the error as $\sigma = \sqrt{\frac{\sigma_+^2+\sigma_-^2}{2}}$, where $\sigma_+$ and $\sigma_-$ denote the $1\sigma$ deviations of upper and lower limits, respectively. In view of the best-fit value and the error of the parameter in the fit, we can calculate the constraint precision of the parameter. For a parameter $\xi$, we can define the constraint precision as $\varepsilon(\xi) = \sigma(\xi)/\xi_{bf}$, in which $\xi_{bf}$  is the best-fit value of $\xi$. We thus list the constraints precision of parameters in the HDE and RDE models for the fits to current only and current+SL 30-year data in Table ~\ref{tab3}.

In Tables ~\ref{tab2} and ~\ref{tab3}, we can clearly see that the precision of parameters is enhanced evidently when the SL 30-year data are combined. Concretely speaking, for the HDE model, the precision of $\Omega_m$, $h$, and $c$ are promoted from 3.55\%, 1.73\%, and 6.89\% to 0.65\%, 0.58\%, and 5.62\%, respectively. In the RDE model, the constraints precision of $\Omega_m$, $h$, and $\alpha$ are improved from 3.64\%, 1.28\%, and 3.94\% to 2.60\%, 1.02\%, and 0.87\%, respectively. The improvements are also fairly remarkable. Therefore, we can conclude that the joint geometric constraints on dark energy models would be improved enormously when a 30-year observation of the SL test is included.

From Figures~\ref{fig1} and~\ref{fig2}, for the two holographic dark energy models in this work, we can find that the SL test can effectively break the existing parameter degeneracies and obviously improve the precision of parameter estimation. The results are consistent with those of previous studies on dark energy models~\cite{msl1,msl2,msl3,msl4}. Hence, we can further confirm that the improvement of the parameter estimation by SL test data should be independent of the cosmological models in the background, which shows that the involvement of SL test in future cosmological constraints is expected to be significant and necessary.

In this paper, we have used the specific best-fit dark energy model as the fiducial model, instead of the $\Lambda$CDM model, to produce the simulated SL test data. The results have shown that this method is very useful to make a clear analysis for the data comparison, i.e., how the SL test breaks the degeneracy (see Figures \ref{fig1} and~\ref{fig2}). For the issue of quantifying the impact of the SL test data on dark energy constraints in the future geometric measurements, such as the space-based project WFIRST, we refer the interested reader to Refs. \cite{msl2,msl3}.

\section{summary}\label{sec:conclu}

In this paper, we have discussed how the redshift drift measurements impact the parameter estimation for the HDE and RDE models. By detecting redshift drift in the spectra of Lyman-$\alpha$ forest of QSOs, the SL test directly measure the expansion rate of the universe in the ``redshift desert" of $2\lesssim z\lesssim5$, which is not covered by any other existing probes. Thus, this method would provide an important supplement to other geometric measurements and be of great significance for cosmology.

Followed by the previous works, in order to guarantee that the simulated SL test data are consistent with other geometric measurement data, we used the best-fit dark energy models constrained by the current combined geometric measurement data as the fiducial models to produce the mock SL test data and then took advantage of these simulated data to do the analysis.

With the help of the SL test data, we showed that, the existing parameter degeneracies would be broken evidently. Compared to the current SN+BAO+CMB+$\rm H_0$ constraint results, we found that the 30-year observation of SL test could enormously improve the constraints for all the considered dark energy models. As for the HDE model, the precisions of $\Omega_m$, $h$, and $c$ based only on current data are constrained to the 3.55\%, 1.73\%, and 6.89\% level, whereas those in light of current+SL 30-year data are constrained to the 0.65\%, 0.58\%, and 5.62\% level, respectively. With regard to the RDE model, the precisions of $\Omega_m$, $h$, and $\alpha$ based only on current data are constrained to the 3.64\%, 1.28\%, and 3.94\% level, whereas those on the basis of current+SL 30-year data are constrained to the 1.02\%, 0.87\%, and 2.60\% level, respectively. Thus, the precisions of parameter constraint can be promoted effectively for the considered dark energy models. The results are consistent with the other extensively studied dark energy models. It can be concluded that the improvement of the constraint precision by SL test data should be independent of the cosmological models in the background. To make this conclusion more convincing, more dark energy models and MG models should be explored in depth.


\acknowledgments

We acknowledge the use of {\tt CosmoMC}.
This work was supported by the Top-Notch Young Talents Program of China, the National Natural Science Foundation of
China under Grants No.~11175042 and No.~11522540, the Provincial Department of Education of
Liaoning under Grant No.~L2012087, and the Fundamental Research Funds for the
Central Universities under Grants No.~N140505002, No.~N140506002, and No.~N140504007.


\end{document}